# A quantum logic gate between a solid-state quantum bit and a photon

Hyochul Kim[1], Ranojoy Bose[1], Thomas C. Shen[1], Glenn S. Solomon[2] & Edo Waks[1, 2,*]

(1)   Department of Electrical and Computer Engineering, IREAP, University of Maryland, College Park and Joint Quantum Institute, University of Maryland, College Park, Maryland 20742, USA

(2)   Joint Quantum Institute, National Institute of Standards and Technology, and University of Maryland, Gaithersburg, Maryland 20899, USA

*       edowaks@umd.edu

**Integrated quantum photonics provides a promising route towards scalable solid-state implementations of quantum networks[1], quantum computers[2,3], and ultra-low power opto-electronic devices[4,5]. A key component for many of these applications is the photonic quantum logic gate, where the quantum state of a solid-state quantum bit (qubit) conditionally controls the state of a photonic qubit. These gates are crucial for development of robust quantum networks[6-8], non-destructive quantum measurements[9,10], and strong photon-photon interactions[11]. Here we experimentally realize a quantum logic gate between an optical photon and a solid-state qubit. The qubit is composed of a quantum dot (QD) strongly coupled to a nano-cavity, which acts as a coherently controllable qubit system that conditionally flips the polarization of a photon on picosecond timescales, implementing a controlled-NOT (cNOT) gate. Our results represent an important step towards solid-state quantum networks and provide a versatile approach for probing QD-photon interactions on ultra-fast timescales.**

QDs are robust and spectrally narrow quantum emitters that have attracted significant interest as solid-state qubits. Various approaches have been pursued for storage and manipulation of quantum information in QDs. One approach has been to exploit neutral exciton transitions that can be controlled all-optically to enable both single qubit operations as well as two-qubit operations between distinguishable excitons in a QD[12]. More recently, major progress has been achieved in coherently manipulating highly stable spin states of a charged QD, which promise significantly longer coherence times[13-17].

Another important property of QDs is that they can be coupled to optical nano-cavities in the strong coupling regime[18-21] where a QD can modify the cavity spectral response[22,23], enabling novel applications such as ultra-fast low photon number optical switching[24-26] and single QD lasing[27]. Furthermore, the strong coupling regime can be exploited to interface these solid-state qubits with a flying photonic qubit through direct QD-photon quantum logic operations, as proposed in a number of theoretical works[6-8]. In order to realize this capability, three essential requirements must be met. First, the QD must possess two quantum states whose coherence time is long compared to the interaction time with the photonic qubit. Second, the qubit states of the QD must be coherently controllable. Finally, the qubit state of the QD must have a strong effect on the quantum state of the photon. Achieving these requirements in a solid-state photonic platform has remained an outstanding challenge.

In this letter we demonstrate that a QD strongly coupled to an optical nanocavity can satisfy all of the above requirements, implementing a solid-state qubit in a cavity system that can perform quantum gates on a photon at picosecond timescales. We experimentally demonstrate a cNOT logic gate between the QD and a photonic qubit, which is a universal quantum operation that can serve as a general light-matter interface for remote entanglements and distributed quantum computation. Our device is composed of an indium arsenide (InAs) QD strongly coupled to a photonic crystal cavity. Fig. 1a illustrates the level structure of an InAs QD, which includes a ground state ($|g\rangle$) and two bright exciton states, labelled $|+\rangle$ and $|-\rangle$, representing the two anti-aligned spin configurations of the electron and hole. The optical transitions from the ground state to the two bright excitons, denoted $\sigma_+$ and $\sigma_-$, exhibit right and left circularly polarized emission respectively at high magnetic field. For all measurements performed in this work the biexciton transition is significantly detuned and can therefore be ignored, enabling the QD to be treated as a three-level system. By applying a magnetic field in the sample growth direction (Faraday configuration), the $\sigma_+$ transition can be tuned on resonance with the cavity while the $\sigma_-$ transition remains detuned[28]. In this configuration, states $|g\rangle$ and $|-\rangle$ are the qubit states of the QD, while the $\sigma_+$ transition is used to couple the qubit to a photon. The cavity serves the dual role of creating a photonic interface through cavity reflectivity modification[22,23] via the $\sigma_+$ transition, and suppressing the spontaneous emission of the $\sigma_-$ transition to timescales that are long compared to the QD-photon interaction time.

Fig. 1b shows a scanning electron micrograph (SEM) image of the fabricated device used to implement a qubit-photon gate, which is composed of a photonic crystal defect cavity coupled to an InAs QD (see Methods and Supplementary section 1 for details on device design and fabrication). Quantum interactions between the QD and a photon are achieved by utilizing the strong dependence of the photonic crystal cavity reflection coefficient on the qubit state of the QD[22,23]. Photonic crystal cavities exhibit high-$Q$ modes that have a well-defined polarization. The photonic qubit encodes quantum information using the polarization states $|H\rangle$ and $|V\rangle$ which are rotated 45° relative to the polarization axis of the cavity. These qubit states can be expressed in the polarization basis that is parallel and orthogonal to the cavity axis, denoted $|x\rangle$ and $|y\rangle$ respectively, using the relations $|H\rangle = (|x\rangle + |y\rangle)/\sqrt{2}$ and $|V\rangle = (|y\rangle - |x\rangle)/\sqrt{2}$. Upon reflection from the sample surface, the photonic qubit states will be transformed to the states $|H\rangle \to (r|x\rangle + |y\rangle)/\sqrt{2}$ and $|V\rangle \to (|y\rangle - r|x\rangle)/\sqrt{2}$ where $r$ is the cavity reflection coefficient. This reflection coefficient can be directly calculated from the Heisenberg-Langevin equations of motion (see Supplementary section 2). If the photon is resonant with the cavity mode and the QD is in state $|-\rangle$ (Fig. 1c bottom), the system behaves like a bare cavity and $r = -1$. The photonic qubit therefore experiences a bit flip ($|H\rangle \to |V\rangle$ and $|V\rangle \to |H\rangle$). If, however, the QD is in state $|g\rangle$ (Fig. 1c top), the optical transition to the $|+\rangle$ state will strongly modify the reflection coefficient. For the special case where both the photon and the $\sigma_+$ transition are resonant with the cavity, the reflection coefficient becomes $r = (C-1)/(C+1)$, where $C = 2g^2/\gamma\kappa$ is the atomic cooperativity. The parameters $g$, $\kappa$ and $\gamma$ represent the cavity-QD coupling strength, cavity energy decay rate, and exciton decay rate for the $\sigma_+$ transition respectively. In the limit that $C \gg 1$, which is expected in the strong coupling regime, we have $r \to 1$ and therefore the photonic qubit remains unchanged ($|H\rangle \to |H\rangle$ and $|V\rangle \to |V\rangle$). Thus, the state of the QD determines whether the photonic qubit will experience a bit flip, which implements a complete cNOT logic gate.

The fabricated device was initially characterized under continuous wave (cw) excitation where the cavity spectrum was measured using a broadband LED as a white light source (see Methods). Fig. 2a plots the cavity reflection spectrum as a function of magnetic field. The spectrum was attained using a cross-polarization configuration where the input field was vertically polarized and the reflected field was analysed in the horizontal direction. At 0 T, the spectrum shows a bright peak due to the cavity along with a second peak due to the QD that is blue detuned from the cavity resonance by 0.11 nm. As the magnetic field is increased, the QD line splits into two peaks corresponding to $\sigma_+$ transition (red shift) and $\sigma_-$ transition (blue shift). As the $\sigma_+$ transition is tuned through the cavity resonance, a clear anti-crossing between the cavity and QD line is observed, which is an indication of strong coupling. Fig. 2b is a high spectral resolution measurement performed using a tunable narrowband laser at 1.6 T (see Methods), when the $\sigma_+$ transition is resonant with the cavity, along with a numerical fit to a theoretical model (see Supplementary Section 3). From the numerical fit we determine $g/2\pi$=12.9 GHz and $\kappa/2\pi$=31.9 GHz ($Q$=10,200). The measured values of $g$ and $\kappa$ satisfy the strong coupling condition $g > \kappa/4$, demonstrating that the device operates in the strong coupling regime[18,19,22].

To populate the $|-\rangle$ state, a tunable narrowband laser was used to excite the sample while simultaneously probing the cavity spectrum with the broadband LED. Fig. 2c shows the broadband LED spectrum as a function of detuning between the pump laser and the $\sigma_-$ transition ($\Delta_L/2\pi$) using a pump power of 1.8 μW (measured before the objective lens). A clear modification of the spectrum is observed when the pump laser becomes resonant with the $\sigma_-$ transition. Figures 2d-f plot the measured spectrum for the specific laser detunings of 10, 0, and -10 GHz respectively. At 0 GHz detuning, the central dip in the cavity spectrum is suppressed due to incoherent pumping of the QD into the $|-\rangle$ state where the QD is decoupled from the cavity mode. This suppression quickly vanishes at both red and blue detuned pump wavelengths.

In order to demonstrate quantum gate operation we utilize short optical pulses to coherently prepare the initial qubit state of the QD as well as to generate the photonic qubit. The lifetime of the |-⟩ state was measured to range from 230 ps to 460 ps depending on cavity detuning (see Supplementary Section 4). Experiments were performed using a 10 ps pump pulse and a 75 ps probe pulse, which are short compared to this timescale. The pump pulse was used to induce Rabi oscillation on the σ_ transition in order to coherently prepare the initial qubit state of the QD, while the weak probe served as the photonic qubit (see Methods). The probe pulse duration was selected to ensure that its spectrum was narrower than the spectral dip in Fig. 2b.

Measurements were first performed by setting the incident probe polarization to be vertically polarized, and measuring the reflected probe intensity along the horizontal polarization axis. Fig. 3a plots the reflected probe intensity as a function of $\sqrt{P}$, where $P$ is the average pump power. The blue circles plot the intensity for an 80 ps pump-probe delay, while the red squares show measurements for a pump-probe delay of 4 ns which is much longer than the lifetime of the |-⟩ state. As the pump power is increased, a clear oscillatory behaviour is observed for 80 ps delay. This sinusoidal behaviour is attributed to Rabi oscillation of the QD between the ground state and the |-⟩ state. The $\pi$ pulse condition is achieved at an average pump power of 0.12 μW. In contrast, no oscillation is observed when the delay is set to 4 ns because the QD has had sufficient time to decay to the ground state after it was excited. The contrast of the Rabi oscillations is observed to decrease with pump intensity, which is in agreement with previous measurements on single excitons, and has been attributed to phonon mediated excitation induced dephasing[29,30].

The full time-resolved reflection spectrum was obtained by tuning the probe beam frequency across the cavity resonance. Figures 3b-e show the measured probe intensity for the 0, $\pi$, $2\pi$, and $3\pi$ pump pulse condition respectively for both 80 ps and 4 ns delay. The measured spectrum for 80 ps delay oscillates from the bare cavity to the cavity-QD coupled spectrum depending on the pump power, demonstrating full control of the cavity reflectivity by coherent manipulation of the qubit state. The relative change in intensity induced by changing the QD from the $|-\rangle$ state (80 ps delay) to the $|g\rangle$ state (4 ns delay), when the probe is on cavity resonance and the pump is at the $\pi$ pulse condition (Fig. 3c), is $(I_{max}-I_{min})/I_{max}=60\pm2\%$. The reduction in contrast in comparison to Fig. 2b is attributed to the finite bandwidth of the probe pulse which is measured to be 4.2 GHz using a narrowband Fabry-Perot cavity filter. Solid lines represent theoretical fits (see Supplementary section 5). The blue curves in Fig. 3c and 3e represent the ideal bare cavity spectra when the QD is excited to the $|-\rangle$ state with unity probability. For 80 ps delay in Fig. 3c, the measured signal at cavity resonant wavelength (920.97 nm) achieves 95% of the maximum predicted value denoted by the blue curve. From this value, the probability of the QD being excited to the $|-\rangle$ state after a $\pi$ pulse is calculated to be $0.93\pm0.04$ (see Supplementary section 5). The small reduction from unity probability is attributed to spontaneous decay of $|-\rangle$ state that may occur before the photonic qubit has finished interacting with the cavity.

To map out the complete relation between the photon polarization and the QD qubit state, Fig. 4 shows pump-probe measurements performed for the four possible combinations of input and output photon polarizations (see Methods). The probe beam frequency was tuned over the cavity resonance while pumping the $\sigma_-$ transition of the QD with a $\pi$-pulse. The pump-probe delay was set to either 80 ps or 4 ns, which correspond to the cases where the QD is in state $|-\rangle$ or $|g\rangle$ respectively. Fig. 4a plots measurement results taken under identical conditions used to obtain Fig. 3c, with the exception that the probe intensity was measured in the vertical polarization direction. In this case, the conjugate effect is observed. When the QD is in state $|-\rangle$ (80 ps delay), the bare cavity spectrum is observed as an anti-resonance instead of a resonance. Similarly, when the QD is in state $|g\rangle$ (4 ns delay), we observe the conjugate cavity-QD coupled spectrum where the measured intensity exhibits a peak at the QD resonant frequency instead of a dip. This conjugate behaviour indicates that when more light is transmitted through the PBS less light is reflected and vice versa. Fig. 4b-d plot the other combinations of input and output polarization.

Optimal gate operation is attained when the input field is resonant with the QD, which occurs at a wavelength of 920.96 nm as indicated by the blue vertical line in Fig. 4a. We calculate the probability table for the quantum gate at this operating condition, which is shown in Fig. 4e. Details of the calculations are provided in Supplementary section 6. When the QD is in state $|-\rangle$, the probability of a bit flip is given by $P_{H \rightarrow V}=0.93 \pm 0.03$ and $P_{V \rightarrow H}=0.98 \pm 0.04$, which give the gate fidelity (the probability of being in the correct output state) for the two input polarizations. The small reduction from ideal gate operation is attributed to spontaneous emission of $|-\rangle$ state as previously discussed. When the QD is in state $|g\rangle$, the gate fidelities are given by $P_{V \rightarrow V}=0.58 \pm 0.04$ and $P_{H \rightarrow H}=0.61 \pm 0.07$. The reduction in gate fidelity in this case is due to finite cooperativity and spectral wandering, as consistent with the contrast measured in Fig. 2b under monochromatic excitation.

In conclusion, we have shown that a solid-state qubit composed of a QD strongly coupled to an optical cavity can conditionally flip the polarization of a photon on picosecond timescales. This operation implements a cNOT gate, an important enabler for robust and scalable quantum networks[6-8]. A cPHASE gate can also be attained by orienting the incident photon polarization parallel to the cavity axis, instead of 45°, providing a method for creating strong photon-photon interactions[11]. Improved switching contrast and greater gate speed could be attained by utilizing photonic crystal cavity designs with smaller mode volumes[31], and by performing better alignment of the QD with the high field mode of the cavity[21]. The method demonstrated in this work can also be extended to solid-state qubits that utilize electron and hole spins of charged QDs, which exhibit significantly longer coherence time[13-17]. The current device implementation could further be transitioned to a planar integrated architecture by using a waveguide coupled cavity-QD system in the strong coupling regime[32]. Incorporation of local tuning methods such as the quantum confined Stark effect could further enable resonant cavity coupling of multiple QDs in an integrated device[33], providing a potential photonic platform for development of quantum information processors on a chip.

**Methods**

**Device fabrication:** The sample consisted of 160 nm GaAs layer on top of 1 μm aluminium gallium arsenide (AlGaAs) sacrificial layer grown by molecular beam epitaxy. A single layer of self-assembled InAs QDs (density of 10-50/μm$^2$) was grown in the center of the GaAs layer. A DBR mirror composed of 10 layers of GaAs and aluminium arsenide (AlAs) was grown below the photonic crystal layer and acted as a high reflectivity mirror, enabling the device to behave as a one sided cavity[22]. Photonic crystal cavities with a three-hole defect (L3 type cavity) were fabricated using electron beam lithography, followed by Cl$_2$ based dry etching, and finally wet etch removal of the AlGaAs sacrificial layer using hydrofluoric acid.

**Measurement Setup:** The sample was mounted in a continuous flow liquid helium cryostat and cooled down to a temperature of 4.3 K. The sample mount was surrounded by a superconducting magnet that can apply magnetic fields of up to 7 T. Sample excitation and collection was performed by confocal microscopy using an objective lens with numerical aperture of 0.68. The polarization axis for excitation and collection was set by a HWP and analysed by a PBS, as illustrated in Fig. 1d. Collected signal was focused onto a single mode fiber to spatially filter only the cavity-coupled signal and isolate a single transverse mode, and then measured by a grating spectrometer and nitrogen cooled CCD camera. The resolution of the spectrometer camera system was determined to be 7 GHz.

**CW measurement:** The cavity spectrum was measured using either a broadband LED or a tunable diode laser. The LED was used as a white light source with dominant emission in the wavelength range of 900~950 nm. The diode laser had a narrow linewidth (< 300 kHz) that could be continuously tuned between 920 and 940 nm. The high resolution cavity spectrum in Fig. 2b was measured by continuously sweeping the tunable laser frequency over the cavity resonance and measuring the reflected laser signal. Each data point in Fig. 2b was obtained by fitting the measured laser signal with a Gaussian function where the frequency and scattering intensity of each data point was obtained from the Gaussian fit. Figures 2c-f were obtained by sweeping a diode laser frequency over the $\sigma_-$ transition to pump the $|-\rangle$ state, while simultaneously probing the cavity spectrum with the broadband LED. Background noise due to inelastic scattering from the pump was subtracted in Figs. 2d-f. The contrast of the dip induced by the QD in Fig. 2d and 2f was measured to be 25% on resonance with the $\sigma_+$ transition, which was lower than the measured contrast in Fig. 2b due to limited spectrometer resolution as well as off-resonant excitation of the $\sigma_+$ transition by the pump laser.

**Pump-probe pulse measurement:** The pump and probe were generated using two time synchronized Ti:Sapphire lasers. The sample was maintained at 4.3 K, and a magnetic field of 3-5 T was applied depending on the detuning between the QD and cavity. The lasers were synchronized by a piezo feedback in the probe laser cavity which locked its clock frequency to the pump laser with an accuracy of 100 fs. The delay between the pump and probe was controlled electronically by a phase-lock loop in the synchronization electronics. The pump pulse duration, initially 2 ps, was expanded to 10 ps by spectral filtering and the probe pulse duration, initially 5 ps, was filtered to 75 ps using separate grating spectrometers. After filtering, the probe beam passed through an intensity stabilizer to keep the intensity constant. The pump-probe delay was measured by single photon avalanche photodiode with 30 ps resolution. The probe beam power was set to 1 nW measured before the objective lens. The coupling efficiency of the probe into the cavity mode was previously measured to be 0.16%[34]. This efficiency, along with the laser repetition rate of 76 MHz indicates that the mean number of probe photons per pulse coupled to the cavity is 0.1. In addition to the probe signal detected in the CCD, an inelastic scattering of the pump was observed. This background, measured to be about 5% at the $\pi$ pump pulse condition and increased to 14% at $3\pi$ pump condition, was subtracted in Fig. 3.

**Measurement of complete input-output relation for photon polarizations:** A HWP placed between the PBS and the objective lens was used to rotate the input photon polarization to either *H* or *V*. After reflection the photon underwent a second pass through the HWP due to the optical configuration of the setup. When the HWP was oriented at 0 °, a detection event from the transmitted port of the PBS corresponded to a photon polarized in the *H* direction after reflection, while a detection event from the reflected port corresponded to *V* polarization. In contrast, when the HWP was rotated 45 ° a detection event at the transmission port of the PBS corresponded to a *V* polarized photon after reflection from the cavity, while the reflection port corresponded to an *H* polarized photon. This additional rotation was taken into account in the data in Fig. 4a and 4d.

**Acknowledgments:** The authors would like to acknowledge support from the ARO MURI on Hybrid quantum interactions (grant number W911NF09104), the Physics Frontier Center at the Joint Quantum Institute, and the ONR Applied Electromagnetics center. E. W. would like to acknowledge support from an NSF CAREER award (grant number ECCS. 0846494) and a DARPA Young Faculty Award (grant number N660011114121).

**Author Contributions:** H.K. and E.W. conceived and designed the experiment, and prepared the manuscript. H.K. carried out the measurement and analyzed the data. R.B. and T.C.S. contributed to the measurement and sample design. E.W. and H.K. carried out the theoretical analysis. G.S.S. provided samples grown by molecular beam epitaxy.

**Author Information:** The authors declare no competing financial interests. Correspondence and requests for materials should be addressed to edowaks@umd.edu.

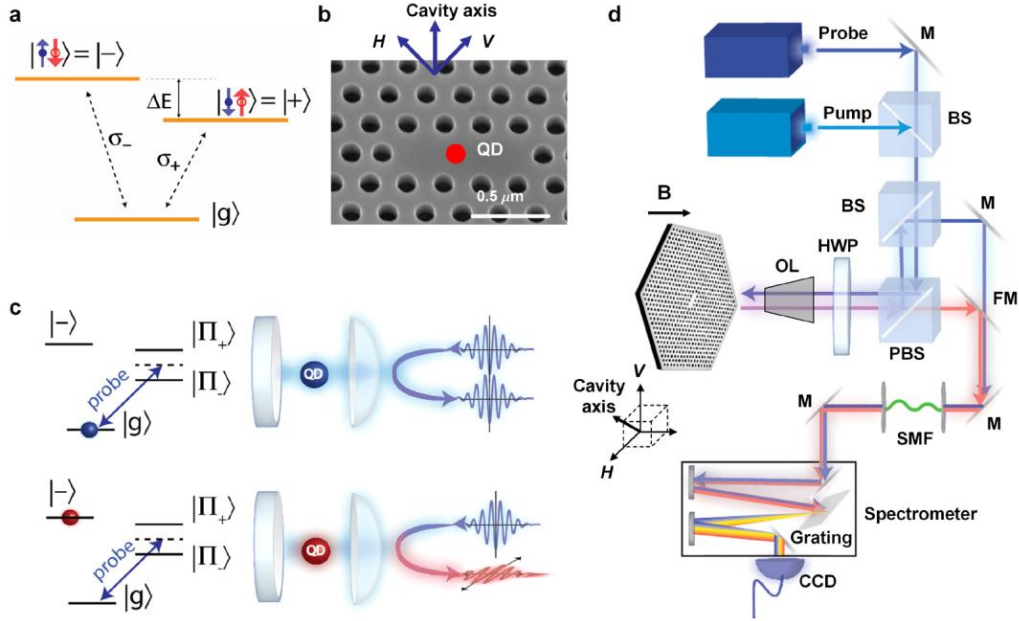

**Figure 1** Implementation of a QD-photon cNOT operation. **a**. Energy level structure of a neutral QD under a magnetic field. **b.** SEM image of the fabricated device and the cavity axis relative to the photon polarization. **c**. Illustration of cNOT operation: The polarization of an incident photon is preserved when the QD is in state $|g\rangle$ (top), and is rotated when the QD is in state $|-\rangle$ (bottom). The horizontal dashed line indicates the degenerate energy level of $|+\rangle$ QD state and the cavity photon state, which are split into two polariton states $|\Pi_+\rangle$ and $|\Pi_-\rangle$ in the strong coupling regime. **d**. Measurement setup. Pump and probe polarization is selected and measured using a polarizing beam splitter (PBS) and a half-wave plate (HWP). A flip mirror (FM) is used to direct the probe signal from either the transmitted or reflected port of the PBS to a single mode fiber (SMF) and then to a grating spectrometer. OL: objective lens, BS: beam splitter and M: mirror.

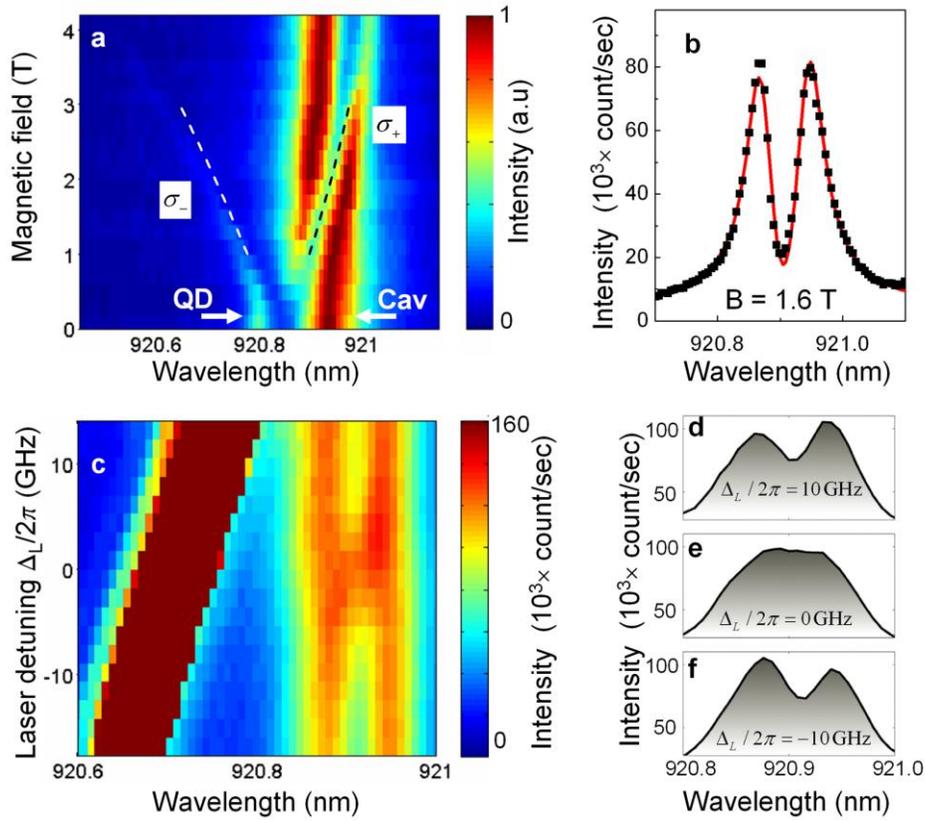

**Figure 2** Device characterizations under cw excitation. **a**. Cavity spectrum measured by a broadband LED as a function of magnetic field at temperature of 4.3 K. **b**. Cavity spectrum measured by a tunable narrowband laser diode at 1.6 T of magnetic field. The red solid line is a fit to a theoretical model. **c**. Cavity spectrum measured by a broadband LED as a function of diode pump laser frequency which is swept across the $\sigma_-$ transition of the QD at magnetic field of 1.6 T. When the pump laser is resonant with the $\sigma_-$ transition, the dip induced by the QD is inhibited. **d-f**. Cavity spectrum for pump laser detuning of $\Delta_L/2\pi$ = 10, 0 and -10 GHz, respectively relative to the $\sigma_-$ transition.

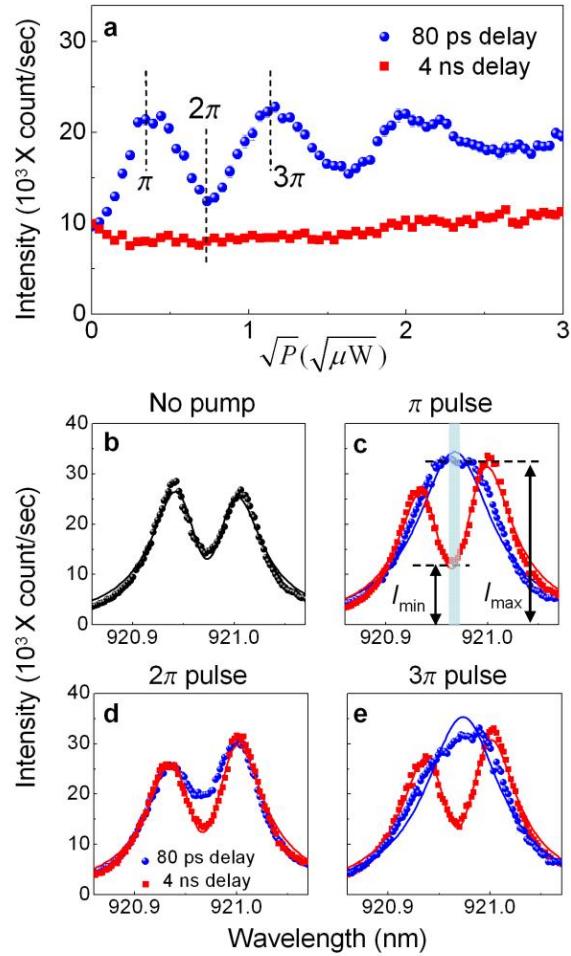

**Figure 3** Demonstration of controlled bit flip by pulsed pump-probe excitation. **a**. Blue circles (red squares) plot the change in measured intensity of the probe signal along the *H* polarization direction as a function of $\sqrt{P}$ at 80 ps (4 ns) pump-probe delay time. **b-e**. Probe signal intensity (*H* polarized) as a function of excitation wavelength at 0, $\pi$, $2\pi$ and $3\pi$ pumping conditions. Blue circles: 80 ps delay, red squares: 4 ns delay. Solid lines are fits to a theoretical model.

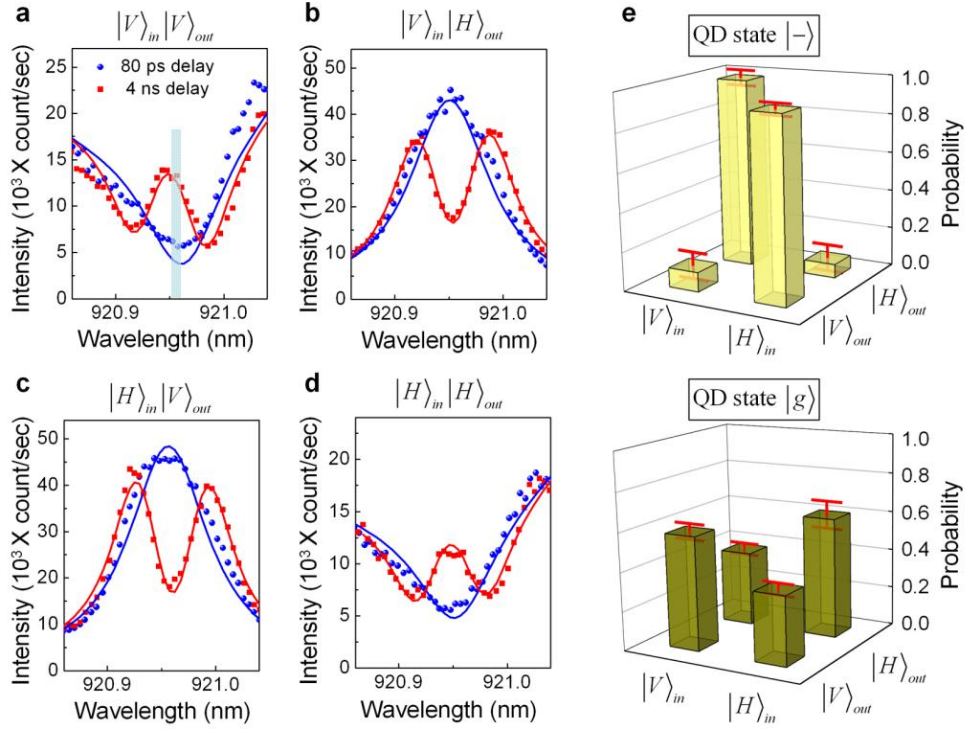

**Figure 4** CNOT operations for all four combinations of input-output polarizations. **a-d**. Cavity spectra are measured with a π pump pulse for 80 ps (blue circles) and 4 ns (red squares) pump-probe delay for four possible combinations of input polarization $|a\rangle_{in}$ and measured polarization $|b\rangle_{out}$ where $a,b \in [\,H, V\,]$. Solid lines are fits to a theoretical model. **e**. Measured probability $P_{a \to b}$ when QD is pumped to state $|\text{-}\rangle$ by a π-pulse (top) and when it has relaxed back to state $|g\rangle$ (bottom).

# Supplementary Material: A quantum logic gate between a solid-state quantum bit and a photon

Hyochul Kim[1], Ranojoy Bose[1], Thomas C. Shen[1], Glenn S. Solomon[2] & Edo Waks[1, 2,*]

(1) Department of Electrical and Computer Engineering, IREAP, University of Maryland, College Park and Joint Quantum Institute, University of Maryland, College Park, Maryland 20742, USA

(2) Joint Quantum Institute, National Institute of Standards and Technology, and University of Maryland, Gaithersburg, Maryland 20899, USA

\* edowaks@umd.edu

## 1. Supplementary Methods

**Details of device design and fabrication**: The photonic crystal cavity is composed of a three-hole defect (L3) cavity as shown in Supplementary Fig. S1. The lattice constant of the photonic crystal was set to $a$=240 nm, the hole radius was set to $r$=70 nm, and the membrane thickness of the gallium arsenide (GaAs) slab was 160 nm. The positions of holes A, B, and C indicated in Fig. S1 were shifted by 42 nm, 6 nm, and 42 nm to optimize the cavity quality factor[1,2]. The cavity decay rate was measured to be $\kappa/2\pi$=31.9 GHz, which corresponds to a cavity quality factor of 10,200. The cavity mode volume was calculated using finite-difference time-domain (FDTD) simulations to be V=0.8 ($\lambda_{cav}/n$)$^3$, where $\lambda_{cav}$ is the wavelength of the cavity resonance and $n=3.6$ is the refractive index of GaAs.

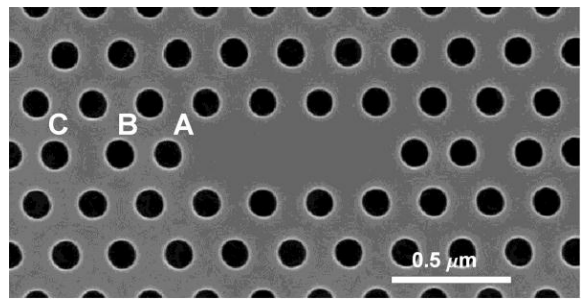

**Supplementary Figure S1.** Scanning electron micrograph image of a GaAs L3 photonic crystal cavity.

Photonic crystal cavities were patterned on a GaAs membrane that contained a single layer of indium arsenide (InAs) quantum dots (QDs) with a QD density of 10-50/μm$^2$. The ground state emission of the QDs varied from approximately 900-950 nm. Considering the small mode volume of the cavities, a strongly coupled QD was found in roughly 5-10% of the devices. The spontaneous emission lifetime and the coherence time of InAs QDs at 4 K has been reported in a number of previous works[3-5] to be around 500-1,000 ps and 400-600 ps, respectively.

**Second order correlation measurement:** A second order correlation measurement was performed to verify that we are working with a single QD. A pulse laser with the repetition rate of 76 MHZ was used to excite the cavity resonantly while the QD was detuned by 0.8 nm. The pulsed laser excited the QD through a phonon mediated non-resonant energy transfer[6]. The QD emission was filtered by a spectrometer grating and sent to the Hanbury Brown-Twiss (HBT) setup composed of a 50/50 beamsplitter and two avalanche photodiodes. Detection events from the counters were processed using a time interval analyser to obtain the correlation measurement. The results of the second order correlation measurement are shown in Supplementary Fig. S2, which plots $g^2(\tau)$ both for a pulsed laser input and for the QD emission. The QD emission shows a nearly complete suppression of the $\tau=0$ case, confirming that the emission is coming from a single QD.

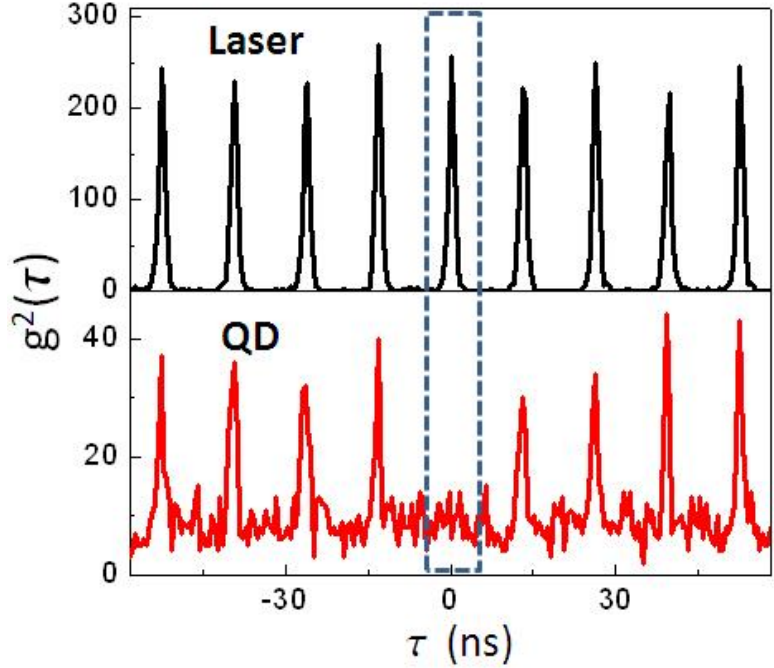

**Supplementary Figure S2.** Second order correlation measurement of the QD emission (bottom) and the excitation pulse laser (top).

## 2. Calculation of cavity reflection coefficient

The cavity reflection coefficient can be directly calculated using cavity input-output formalism[7]. Supplementary Fig. S3 shows the definitions of the cavity input and output field operators, along with the definitions of the polarization axes. Photonic crystal L3 cavities have a well-defined polarization axis along the direction orthogonal to the row defect[1]. We define $\hat{\mathbf{a}}_x$ and $\hat{\mathbf{a}}_y$ as bosonic input flux operators[7] for a photon that is parallel and orthogonal to the polarization axis of the cavity respectively. The flux operators can also be expressed in the *H-V* basis, rotated 45° relative to the cavity polarization axis, via the relations

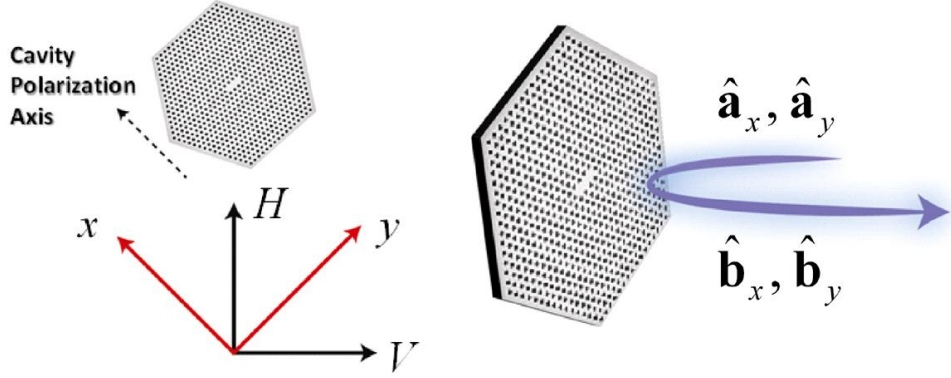

**Supplementary Figure S3.** Definition of polarization angles and input and output modes in theoretical model.

$$\hat{\mathbf{a}}_H = \frac{\hat{\mathbf{a}}_x + \hat{\mathbf{a}}_y}{\sqrt{2}} \quad (1)$$

$$\hat{\mathbf{a}}_V = \frac{\hat{\mathbf{a}}_y - \hat{\mathbf{a}}_x}{\sqrt{2}} \quad (2)$$

The cavity output operators $\hat{\mathbf{b}}_x$ and $\hat{\mathbf{b}}_y$ in the polarization basis of the cavity are related to the input operators by the cavity input-output relations

$$\hat{\mathbf{b}}_x = \hat{\mathbf{a}}_x - \sqrt{\kappa}\hat{\mathbf{a}} \quad (3)$$

$$\hat{\mathbf{b}}_y = \hat{\mathbf{a}}_y \quad (4)$$

where $\hat{\mathbf{a}}$ is the bosonic annihilation operator for a cavity photon. Eq. 4 shows that the reflection coefficient for a photon polarized in the $y$ direction is $r_y = 1$, which is expected because a $y$-polarized photon does not couple to the cavity and simply reflects from the sample surface. The reflection coefficient for a photon polarized along the $x$-axis is more complicated, and depends on the interaction between the QD and the cavity. It has been derived in a number of previous works using various approaches[8-10]. For completeness, we provide a derivation below using Heisenberg-Langevin formalism.

In order to derive the reflection coefficient for an *x*-polarized photon, an expression must be attained for the cavity operator $\hat{\mathbf{a}}$. This expression can be attained from the Heisenberg-Langevin equations of motion for an atomic system coupled to an optical cavity[11]. We derive this expression in the limit that the $\sigma_-$ transition is highly detuned from the cavity mode, and is therefore optically decoupled from the cavity. Thus, the cavity reflection coefficient is dominated by the interaction with the $\sigma_+$ transition. The Hamiltonian for a closed cavity mode coupled to a two-level atom is described by

$$\mathbf{H} = \hbar\omega_c \mathbf{a}^\dagger \mathbf{a} + \frac{\hbar\omega_a}{2}\hat{\mathbf{w}} + \hbar g\left(\mathbf{a}^\dagger\hat{\mathbf{s}} + \hat{\mathbf{s}}^\dagger\mathbf{a}\right) \tag{5}$$

In the above, $\hat{\mathbf{s}} = |g\rangle\langle+|$ is defined as the QD lowering operator, and $\hat{\mathbf{w}} = |+\rangle\langle+| - |g\rangle\langle g|$ is the population difference operator. The frequencies $\omega_c$ and $\omega_a$ are the cavity and QD resonant frequencies respectively, while *g* is the cavity-QD coupling strength. In order to include losses in both the cavity and QD, as well as cavity excitation, we apply the Heisenberg-Langevin formalism[11] to attain the Heisenberg-Langevin equations of motion

$$\frac{d\hat{\mathbf{a}}}{dt} = -\left(i\Delta_c + \frac{\kappa}{2}\right)\hat{\mathbf{a}} - ig\hat{\mathbf{s}} + \sqrt{\kappa}\hat{\mathbf{a}}_x \tag{6}$$

$$\frac{d\hat{\mathbf{s}}}{dt} = -\left(i\Delta_a + \frac{\Gamma_{spon}}{2}\right)\hat{\mathbf{s}} + ig\hat{\mathbf{w}}\hat{\mathbf{a}} \tag{7}$$

$$\frac{d\hat{\mathbf{w}}}{dt} = -\Gamma_{spon}\left(\hat{\mathbf{w}} + \hat{\mathbf{I}}\right) + 2ig\left(\hat{\mathbf{a}}^\dagger\hat{\mathbf{s}} + \hat{\mathbf{s}}^\dagger\hat{\mathbf{a}}\right) \tag{8}$$

where $\kappa$ is the cavity decay rate, $\Gamma_{spon}$ is the QD spontaneous emission rate, and $\hat{\mathbf{a}}_x$ is the cavity input operator defined in Fig. S3 which drives the cavity mode. In the above equations we have assumed that the input field spectrum is centered around a frequency ω, and have transformed the equations of motion to the reference frame rotating at this field frequency. We defined $\Delta_c = \omega_c - \omega$ and $\Delta_a = \omega_a - \omega$. Eq. (8) can be directly integrated to give

$$\hat{\mathbf{w}}(t) = \hat{\mathbf{w}}_0 e^{-\Gamma_{spon} t} + \left(e^{-\Gamma_{spon} t} - 1\right)\hat{\mathbf{I}} + \hat{\mathbf{R}} \tag{9}$$

where

$$\hat{\mathbf{R}} = 2ig \int_0^t \left(\hat{\mathbf{a}}^\dagger \hat{\mathbf{s}} + \hat{\mathbf{s}}^\dagger \hat{\mathbf{a}}\right) dt' \tag{10}$$

To proceed, the initial state of the QD, which can be either $|g\rangle$ or $|-\rangle$, must be specified. These two cases are considered individually below for the case where the input field is monochromatic. The solution will then be extended to non-monochromatic fields so that it can be applied to both pulsed and broadband continuous wave input fields.

**QD is in state $|g\rangle$:** Inserting Eq. (9) into Eq. (7) and taking the expectation of both sides we attain

$$\frac{d\langle\hat{\mathbf{a}}\rangle}{dt} = -\left(i\Delta_c + \frac{\kappa}{2}\right)\langle\hat{\mathbf{a}}\rangle - ig\langle\hat{\mathbf{s}}\rangle + \sqrt{\kappa}\langle\hat{\mathbf{a}}_x\rangle \tag{11}$$

$$\frac{d\langle\hat{\mathbf{s}}\rangle}{dt} = -\left(i\Delta_a + \frac{\Gamma_{spon}}{2}\right)\langle\hat{\mathbf{s}}\rangle - ig\langle\hat{\mathbf{a}}\rangle + \langle\hat{\mathbf{R}}\hat{\mathbf{a}}\rangle \tag{12}$$

In the above equation, we used the fact that the QD is initially in the state $|g\rangle$ which means that $\langle\hat{\mathbf{w}}_0 \hat{\mathbf{O}}\rangle = -\langle\hat{\mathbf{O}}\rangle$ for any operator $\hat{\mathbf{O}}$. We consider the limit that there is only one excitation in the system. This regime, called the weak field limit, is exact if the input field is a single photon state. It also provides an accurate approximation to the cavity response if the cavity is excited by a weak coherent field with an average photon number that is much less than 1. In the weak field limit we have $\langle\hat{\mathbf{R}}\hat{\mathbf{a}}\rangle = 0$. This result can be understood using Eq. (10) which shows that the operator $\hat{\mathbf{R}}\hat{\mathbf{a}}$ annihilates two excitations (an excitation can be either a cavity photon or a QD exciton), and then creates a single excitation. If there is only one excitation in the entire system, this term will always annihilate the initial state and therefore does not contribute.

Within the weak field regime, the dipole moment and average cavity field are related by a system of linear constant coefficient differential equations given by

$$\frac{d\langle \hat{\mathbf{a}} \rangle}{dt} = -\left(i\Delta_c + \frac{\kappa}{2}\right)\langle \hat{\mathbf{a}} \rangle - ig\langle \hat{\mathbf{s}} \rangle + \sqrt{\kappa}\langle \hat{\mathbf{a}}_x \rangle \quad (13)$$

$$\frac{d\langle \hat{\mathbf{s}} \rangle}{dt} = -(i\Delta_a + \gamma)\langle \hat{\mathbf{s}} \rangle - ig\langle \hat{\mathbf{a}} \rangle \quad (14)$$

where $\gamma = \Gamma_{spon}/2 + 1/T_2$ is the QD homogeneous linewidth and $T_2$ is the pure dephasing time, which has been introduced into the mean operator equations using the random phase jump approach[12]. We consider here the case where the mean input field $\langle \hat{\mathbf{a}}_x \rangle$ is a monochromatic field. In this limit one can solve for the mean cavity field by taking the steady-state solution to the above equations. This solution is given by

$$\langle \hat{\mathbf{a}} \rangle = \frac{\sqrt{\kappa}(i\Delta_a + \gamma)}{(i\Delta_c + \kappa/2)(i\Delta_a + \gamma) + g^2}\langle \hat{\mathbf{a}}_x \rangle \quad (15)$$

Taking the expectation of both sides of Eq. (3) and inserting Eq. (15) we attain the relation $\langle \hat{\mathbf{b}}_x(\omega) \rangle = r(\omega)\langle \hat{\mathbf{a}}_x(\omega) \rangle$ where

$$r(\omega) = 1 - \frac{\kappa(i\Delta_a + \gamma)}{(i\Delta_c + \kappa/2)(i\Delta_a + \gamma) + g^2} \quad (16)$$

For the special case where both the QD and field are resonant with the cavity, we have $\Delta_a = \Delta_c = 0$ and the reflection coefficient takes on the simplified expression

$$r(\omega) = \frac{C-1}{C+1} \quad (17)$$

where $C = 2g^2/\gamma\kappa$ is the atomic cooperativity, which is the expression quoted in the main manuscript.

**QD is in state |-⟩:** We consider the limit where the lifetime of state |-⟩ is long compared to the temporal dynamics of all input fields. In this limit the QD will remain in state |-⟩ throughout the entire time that the input field is interacting with the cavity. Thus, $\langle \hat{\mathbf{s}} \rangle = 0$ at all time, and Eq. (13) simplifies to

$$\frac{d\langle\hat{\mathbf{a}}\rangle}{dt} = -\left(i\Delta_c + \frac{\kappa}{2}\right)\langle\hat{\mathbf{a}}\rangle + \sqrt{\kappa}\langle\hat{\mathbf{a}}_x\rangle \tag{18}$$

The above equation is that of a bare cavity driven by an input field, and is exact both in the weak and strong field limit. For a monochromatic field, we can once again take the steady-state solution given by

$$\langle\hat{\mathbf{a}}\rangle = \frac{\sqrt{\kappa}}{i\Delta_c + \kappa/2}\langle\hat{\mathbf{a}}_x\rangle \tag{19}$$

Taking the expectation of both sides of Eq. (3) and inserting the Eq. (19) we attain the relation $\langle\hat{\mathbf{b}}_x(\omega)\rangle = r(\omega)\langle\hat{\mathbf{a}}_x(\omega)\rangle$ where

$$r(\omega) = 1 - \frac{\kappa}{i\Delta_c + \kappa/2} \tag{20}$$

For the special case where the input field is resonant with the cavity ($\Delta_c = 0$), we have $r(\omega) = -1$.

**Conversion of results to *H/V* basis:** The above results can be expressed in the *H-V* basis using the relation

$$\hat{\mathbf{b}}_H = \frac{\hat{\mathbf{b}}_x + \hat{\mathbf{b}}_y}{\sqrt{2}} \tag{21}$$

$$\hat{\mathbf{b}}_V = \frac{\hat{\mathbf{b}}_y - \hat{\mathbf{b}}_x}{\sqrt{2}} \tag{22}$$

Taking the expectation values of the above equations along with the relations $\langle\hat{\mathbf{b}}_y\rangle = \langle\hat{\mathbf{a}}_y\rangle$ and $\langle\hat{\mathbf{b}}_y\rangle = r(\omega)\langle\hat{\mathbf{a}}_y\rangle$ as well as Eq. (1) and Eq. (2) we attain the relations

$$\langle\hat{\mathbf{b}}_H\rangle = \frac{1+r(\omega)}{2}\langle\hat{\mathbf{a}}_H\rangle + \frac{1-r(\omega)}{2}\langle\hat{\mathbf{a}}_V\rangle \tag{23}$$

$$\langle\hat{\mathbf{b}}_V\rangle = \frac{1-r(\omega)}{2}\langle\hat{\mathbf{a}}_H\rangle + \frac{1+r(\omega)}{2}\langle\hat{\mathbf{a}}_V\rangle \tag{24}$$

The above relations enable us to calculate the output field amplitudes as a function of the input field amplitudes along the *H* and *V* axes.

**Extension to non-monochromatic fields:** The previous results were calculated in the limit of a monochromatic input field. This result can be extended to non-monochromatic fields in a straightforward way. In the weak-field limit, the cavity-QD system is described by a system of constant-coefficient differential equations and is therefore a linear system. Thus, for a non-monochromatic field each frequency component will interact with the cavity independently. The response of the field can be directly attained by performing Fourier decomposition of the input field and calculating the reflectivity of each frequency component independently. That is, we can write

$$\langle \hat{\mathbf{a}}_x(t) \rangle = \int \varepsilon_x(\omega) e^{-i\omega t} d\omega \tag{25}$$

where $\varepsilon_x(\omega)$ is the Fourier component of the field amplitude. We then have

$$\langle \hat{\mathbf{b}}_x(t) \rangle = \int r(\omega) \varepsilon_x(\omega) e^{-i\omega t} d\omega \tag{26}$$

Similarly, along the *y* axis we have $r(\omega) = 1$ and hence

$$\langle \hat{\mathbf{b}}_y(t) \rangle = \int \varepsilon_y(\omega) e^{-i\omega t} d\omega \tag{27}$$

The above amplitudes can be expressed in the *H/V* basis using the same approach as for the monochromatic field, which leads to the expressions

$$\langle \hat{\mathbf{b}}_H(t) \rangle = \int \left[ \frac{1+r(\omega)}{2} \varepsilon_H(\omega) + \frac{1-r(\omega)}{2} \varepsilon_V(\omega) \right] e^{-i\omega t} d\omega \tag{28}$$

$$\langle \hat{\mathbf{b}}_V(t) \rangle = \int \left[ \frac{1-r(\omega)}{2} \varepsilon_H(\omega) + \frac{1+r(\omega)}{2} \varepsilon_V(\omega) \right] e^{-i\omega t} d\omega \tag{29}$$

Most of the time, we are primarily interested in the total energy reflected by the cavity (i.e. average number of reflected photons per pulse). This value can be attained using Parseval's theorem which states

$$W_H = \frac{1}{2\pi} \int_{-\infty}^{\infty} \left| \frac{1+r(\omega)}{2} \varepsilon_H(\omega) + \frac{1-r(\omega)}{2} \varepsilon_V(\omega) \right|^2 d\omega \qquad (30)$$

$$W_V = \frac{1}{2\pi} \int_{-\infty}^{\infty} \left| \frac{1-r(\omega)}{2} \varepsilon_H(\omega) + \frac{1+r(\omega)}{2} \varepsilon_V(\omega) \right|^2 d\omega \qquad (31)$$

If the input field is vertically polarized we have $\varepsilon_v(\omega) = 0$ and the above equations simplify to

$$W_{V \to H} = \int_{-\infty}^{\infty} \left| \frac{1-r(\omega)}{2} \right|^2 S_{in}^V(\omega) d\omega \qquad (32)$$

$$W_{V \to V} = \int_{-\infty}^{\infty} \left| \frac{1+r(\omega)}{2} \right|^2 S_{in}^V(\omega) d\omega \qquad (33)$$

where $S_{in}^V(\omega) = |\varepsilon_v(\omega)|^2 / 2\pi$ is the input power spectrum. Similarly, if the input field is horizontally polarized we can write

$$W_{H \to H} = \int_{-\infty}^{\infty} \left| \frac{1+r(\omega)}{2} \right|^2 S_{in}^H(\omega) d\omega \qquad (34)$$

$$W_{H \to V} = \int_{-\infty}^{\infty} \left| \frac{1-r(\omega)}{2} \right|^2 S_{in}^H(\omega) d\omega \qquad (35)$$

where $S_{in}^H(\omega) = |\varepsilon_H(\omega)|^2 / 2\pi$.

### 3. Numerical fit of reflection spectrum

To this point, the QD linewidth was assumed to be homogeneously broadened. Real QDs exhibit spectral diffusion where the QD frequency wanders on timescales that are long compared to the laser repetition rate, but short compared to the total measurement time. Spectral diffusion in a QD coupled to a photonic crystal cavity has been reported in previous work, and may be due to proximity of the QD to surface[13] as well as thermal fluctuation[14]. This spectral diffusion can be modelled by setting $\omega_{QD} = \omega_0 + \beta$ where $\omega_0$ is the average transition frequency of the $\sigma_+$ transition, and $\beta$ is a zero-mean random variable that describes fluctuations in the QD resonant frequency due to spectral diffusion. With this definition $\Delta_a = \omega - \omega_{QD} = \Delta_a^0 - \beta$ becomes a random variable where $\Delta_a^0 = \omega - \omega_0$ is the mean detuning between the QD and the cavity. The reflection coefficient given in Eq. (16) now becomes

$$r(\omega,\beta) = 1 - \frac{\kappa\left[i(\Delta_a^0 + \beta) + \gamma\right]}{\left[i\Delta_c + \kappa/2\right]\left[i(\Delta_a^0 + \beta) + \gamma\right] + g^2} \qquad (36)$$

Calculated results must now be averaged over the possible values of $\beta$. Thus, Eq. (32)-(35) become

$$W_{V \to H} = \int_{-\infty}^{\infty} P(\beta) \int_{-\infty}^{\infty} \left|\frac{1 - r(\omega,\beta)}{2}\right|^2 S_{in}^V(\omega) d\omega d\beta \qquad (37)$$

$$W_{V \to V} = \int_{-\infty}^{\infty} P(\beta) \int_{-\infty}^{\infty} \left|\frac{1 + r(\omega,\beta)}{2}\right|^2 S_{in}^V(\omega) d\omega d\beta \qquad (38)$$

$$W_{H \to H} = \int_{-\infty}^{\infty} P(\beta) \int_{-\infty}^{\infty} \left|\frac{1 + r(\omega,\beta)}{2}\right|^2 S_{in}^H(\omega) d\omega d\beta \qquad (39)$$

$$W_{H \to V} = \int_{-\infty}^{\infty} P(\beta) \int_{-\infty}^{\infty} \left|\frac{1 - r(\omega,\beta)}{2}\right|^2 S_{in}^H(\omega) d\omega d\beta \qquad (40)$$

where $P(\beta)$ is a probability distribution function that characterizes the inhomogeneous linewidth of the QD.

In Fig. 2b of the main text, a cavity reflectivity measurement was performed with a vertically polarized input field with a narrowband tunable laser. In this case, the input field is very close to a monochromatic field with frequency $\omega_f$ so one can substitute $S_{in}^V(\omega) = W_0^V \delta(\omega - \omega_f)$ which results in the relation

$$W_{V \to H} = W_0^V \int_{-\infty}^{\infty} P(\beta) \left| \frac{1 - r(\omega_f, \beta)}{2} \right|^2 d\beta \tag{41}$$

To perform calculations we need an expression for the QD lineshape function $P(\beta)$. Spectral diffusion is often be modelled using a Guassian distribution given by the relation

$$P(\beta) = \frac{1}{\sqrt{2\pi\gamma_I^2}} \exp\left(-\frac{\beta^2}{2\gamma_I^2}\right) \tag{42}$$

where $\gamma_I$ is the inhomogeneous linewidth due to spectral wandering. We use this analytical expression to fit to the experimental data shown in Fig. 2b of the main text. The variable $W_0^V$ is treated as a fitting parameter, along with $g$, $\kappa$, and $\gamma_I$, while we set $\gamma = \Gamma_{spon}/2$ where $\Gamma_{spon} = 1/530$ ps$^{-1}$ attained from the Purcell factor measurements (see Supplementary Section 4 below). This model assumes that the pure dephasing rate of the QD is negligibly small compared to the inhomogeneous linewidth, which is highly realistic for our system. The result of the fit is plotted as a solid line in Fig. 2b. The fitted values correspond to g/2π = 12.9 GHz, κ/2π=31.9 GHz, and γ$_I$/2π = 5.2 GHz. The Gaussian model for the inhomogeneous linewidth provides very good agreement with experimental results.

Theoretical fits for Fig. 3 and 4 were performed using Eq. (37)-(40) using the approximation that $S_{in}^V(\omega) = W_0^V \delta(\omega - \omega_f)$ and $S_{in}^H(\omega) = W_0^H \delta(\omega - \omega_f)$. These approximations are valid because the laser bandwidth is only 4.2 GHz, which is small compared to both $\kappa$ and 2$g$. In addition, for the cases where the reflected field was measured in the same polarization direction as the incident field polarization (Fig. 4a and 4d), a frequency dependent background level was observed. This frequency dependent background was due to imperfect mode matching of the probe to the cavity, along with the dispersive properties of the polarization optics. These non-idealities in the probing system resulted in a portion of direct reflection from the slab surface which did not couple to the cavity. Direct uncoupled reflection is always measured in the same input and output photon polarization direction because it preserves the polarization of the incident probe. To account for this background, a frequency dependent additive background term was added to the fit. The background intensity was expanded to second order in frequency around the cavity resonance as $I_B(\omega) = a_0 + a_1(\omega - \omega_{cav}) + a_2(\omega - \omega_{cav})^2$ where $a_0$, $a_1$, and $a_2$ were treated as fitting parameters.

## 4. Qubit lifetime measurement

The qubit lifetime can be modified through the Purcell effect by controlling the detuning between the $\sigma_-$ transition and the cavity. Supplementary Fig. S4 plots the change in the probe signal intensity as a function of pump-probe delay time for different detunings ($\Delta$) between the $\sigma_-$ transition and the cavity. The pump was set to the $\pi$-pulse power and tuned to resonance with the $\sigma_-$ transition. The detuning was adjusted by utilizing a slow red-shift of the cavity which was observed during the course of the measurements due to gradual condensation of residual gas deposition on the photonic crystal nanocavity at a cryogenic temperature[15,16]. A slow red shift of 0.1~0.3 nm was observed over several hours, which could be recovered after warming up and cooling back down. By taking measurements at different times relative to the initial cooldown, it was possible to adjust the detuning between the QD and the cavity mode. The magnetic field was set for each value of the cavity resonant frequency so that the $\sigma_+$ transition was resonant with the cavity and the probe beam frequency was set to the $\sigma_+$ transition frequency.

All measurements were taken at a temperature of 4.3 K. For the qubit lifetime measurements, a shorter probe pulse of 22 ps was generated using a tunable Fabry-Perot cavity. Shorter probe pulses were utilized in these measurements in order to improve temporal resolution. Supplementary Fig. S4a shows lifetime measurements for three different QD detunings of 113 GHz, 169 GHz, and 230 GHz. Data points represent experimental measurements while the solid lines represent an exponential fit, which was used to determine the lifetime of the $|-\rangle$ state. The measured lifetime of the three detunings are given by 230 ps, 350 ps, and 460 ps respectively. Fig. S4b plots QD lifetime vs QD detuning along with a theoretical fit to the predicted value given by $\Gamma_{\sigma_-} = 4g^2\kappa/(4\Delta^2+\kappa^2)+\Gamma_0$, where $\Gamma_0$ is a fitting parameter that accounts for nonradiative decay as well as radiative decay into leaky modes. From the fit we determine $1/\Gamma_0=530$ ps.

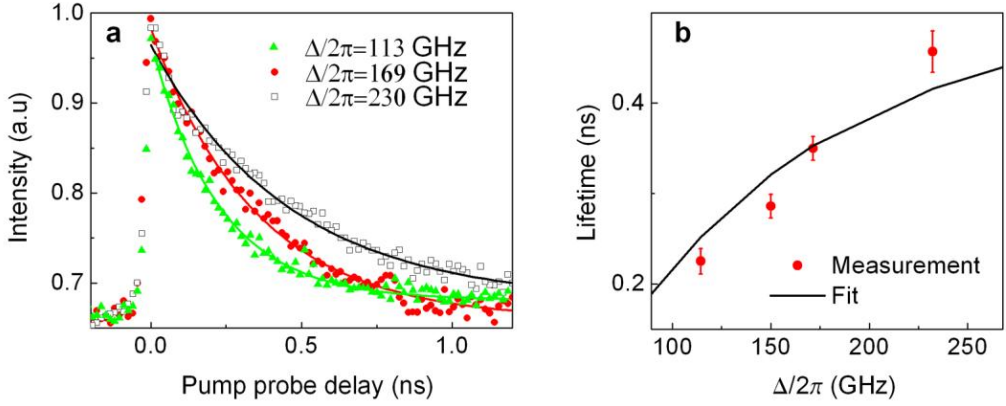

**Supplementary Figure S4.** Qubit lifetime measurement. **a**. Measured probe intensity as a function of pump-probe delay. Solid lines are fits to an exponential decay function. Fitted lifetimes are $1/\Gamma_{\sigma_-}=$ 460 ps for the black curve, 350 ps for the red curve, and 230 ps for the green curve. **b**. Measured (red circles) lifetime as a function of QD ($\sigma_-$)-cavity detuning ($\Delta/2\pi$). A black solid line plots a theoretical fit.

## 5. Probability of QD excitation after a π pulse

In Fig. 3c of the main text, blue circles correspond to measurements taken 80 ps after a π pulse, while red squares are obtained for a 4 ns delay where the QD has fully relaxed back to state |g⟩. The solid red curve represents a numerical fit of Eq. (37) to the data of 4 ns delay, which we define as $W^g_{V \to H}(\lambda)$. We defined $W^-_{V \to H}(\lambda)$ as the ideal intensity distribution when the QD is in state |-⟩ with unity probability throughout the interaction time of the photonic qubit (i.e. when the QD is fully decoupled from the cavity). This distribution is obtained by taking the numerical fit to $W^g_{V \to H}(\lambda)$ and setting g=0, as plotted by the blue curve in Fig. 3c.

Fig. 3c shows that the experimental data for 80 ps delay does not attain the maximum value predicted by $W^-_{V \to H}(\lambda)$. We attribute this degradation to a small probability that the QD has relaxed back down to state |g⟩ before the photonic qubit has finished interacting with the cavity-QD system. We define α as the probability that the QD is in state |-⟩ when the photonic qubit is reflected from the cavity. We then define a new distribution

$W_{V \to H}^{\pi}(\lambda) = \alpha W_{V \to H}^{-}(\lambda) + (1-\alpha) W_{V \to H}^{g}(\lambda)$ which represents a statistical mixture for the case where the QD is in state $|-\rangle$ with probability $\alpha$ and state $|g\rangle$ with probability $1-\alpha$. We treat $\alpha$ as a fitting parameter while fixing all other parameters to values attained from $W_{V \to H}^{g}(\lambda)$ (the red curve), and use this distribution to fit the 80 ps delay data (blue circles). From the fit we obtain $\alpha = 0.93 \pm 0.04$. At $\lambda = 920.97$ nm (the cavity resonant wavelength), we have $W_{V \to H}^{\pi} / W_{V \to H}^{-} = 0.95 \pm 0.03$.

## 6. Calculation of probability table

We use the data plotted in Fig. 4 of the main text to calculate $P_{i \to j}$ where $i,j \in [H,V]$. Supplementary Fig. S5 shows the identical graphs as in the Fig. 4(a) and (c) in the main manuscript, where we have defined the relevant intensities used to calculate the probability table. Fig. S5(a) is the case where the input field is horizontally polarized and the output field is measured in the vertically polarized basis. The solid red curve represents a numerical fit of Eq. (40) to the data of 4 ns delay (red squares), which we define as $W_{H \to V}^{g}(\lambda)$. In order to calculate the probabilities, we need to know the intensity $I_0$, which is the reflected intensity at cavity resonant frequency attained when the QD is in state $|-\rangle$ with unity probability. This value can be attained from the peak value of $W_{H \to V}^{-}(\lambda)$, defined in Supplementary section 5, which gives the ideal intensity distribution when the QD is fully excited to state $|-\rangle$.

Similar to Supplementary section 5, we define $W_{H \to V}^{\pi}(\lambda) = \alpha W_{H \to V}^{-}(\lambda) + (1-\alpha) W_{H \to V}^{g}(\lambda)$ which represents a statistical mixture of distributions for the case where the QD can either be in state $|-\rangle$ or state $|g\rangle$, where $\alpha$ is the probability of the QD being in state $|-\rangle$. This distribution is used to fit the 80 ps delay data (blue circles), and the result is plotted as the black line in Fig. S5(a). The green dashed line plots the distribution $W_{H \to V}^{\infty}(\lambda) = \lim_{g \to \infty} W_{H \to V}^{g}(\lambda)$ which represents the ideal case where the cavity-QD coupling strength is extremely large. This curve does not completely drop to zero because of a small background level, which is about 1% of $I_0$.

Fig. S5(b) shows the case where the input field is vertically polarized and is measured in the vertical polarization basis. The distributions $W^g_{V\to V}(\lambda)$, $W^-_{V\to V}(\lambda)$, $W^\pi_{V\to V}(\lambda)$, and $W^\infty_{V\to V}(\lambda)$ (defined in the identical way to panel a) are plotted as well. We note that $I_0$ does not go all the way down to zero due to the presence of a background intensity level. This background is due to direct uncoupled reflection from the slab surface that is not fully rejected by the single-mode fiber mode filter. The background level indicated as $I_{bg}$ is 19% of ideal cavity scattering signal ($|I_\infty - I_0|$), which is higher than the 1% background observed in Fig. S5(a). The reason for this difference in background levels is that direct surface reflection is observed significantly in the port that measures the same input and output polarization (Fig. S5b) and is not observed in the orthogonal polarization (Fig. S5a) because surface reflection will not rotate the photon polarization. The actual value of the background is strongly depends on the focusing condition onto the cavity and single mode fiber, as well as the optical alignment.

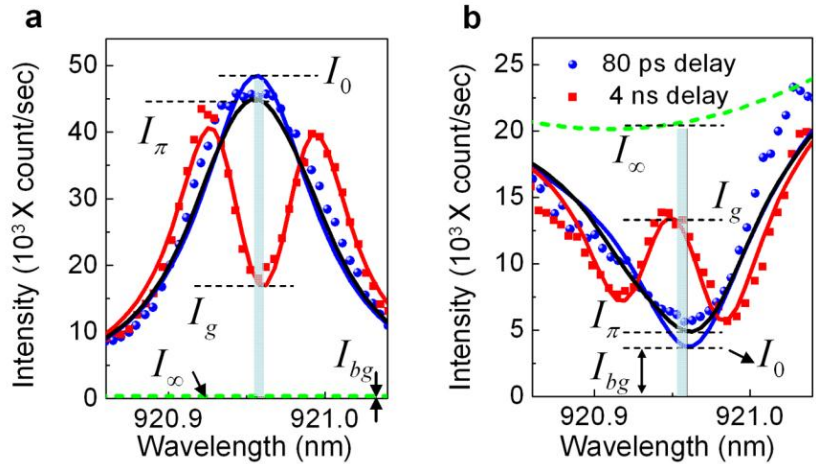

**Supplementary Figure S5.** Duplication of Fig 4(a) and (c) in main manuscript. Cavity spectra are measured with a π pump pulse for 80 ps (blue circles) and 4 ns (red squares) pump-probe delay for (a) input polarization of $|H\rangle_{in}$ and measured polarizations of $|V\rangle_{out}$ and (b) $|V\rangle_{in}$ and $|V\rangle_{out}$. Red curve plots distribution for $W^g_{i\to j}(\lambda)$, blue curve plots the distribution for $W^-_{i\to j}(\lambda)$, black curve plots $W^\pi_{i\to j}(\lambda)$, and green curve plots $W^\infty_{i\to j}(\lambda)$ where $(i,j)=(H,V)$ for panel a and $(i,j)=(V,V)$ for panel b.

We define $P^0_{i\to j}$ as the probability that a photon with initial polarization state $i\in[H,V]$ is detected in the polarization direction $j\in[H,V]$ when the QD is in state $|g\rangle$. Similarly, $P^\pi_{i\to j}$ is defined as the probability for the case where a π pulse has been applied to the QD. The probability is defined as $P^0_{i\to j}=(I_g-I_\infty)/(I_0-I_\infty)$ and $P^\pi_{i\to j}=(I_\pi-I_\infty)/(I_0-I_\infty)$ for the case that input and output photon polarizations are orthogonal (Fig. S5a), and $P^0_{i\to j}=(I_g-I_0)/(I_\infty-I_0)$ and $P^\pi_{i\to j}=(I_\pi-I_0)/(I_\infty-I_0)$ for the case where the input and output photon polarizations are same (Fig. S5b). Here, $I_g$, $I_\pi$, $I_\infty$, and $I_0$ are the reflected intensity attained for the red, black, green and blue curves at cavity resonant wavelength (920.96 nm) as indicated in Fig. S5. A table of all calculated probabilities is shown below. Error values are determined from a 95% confidence bound of the numerical fits.

| QD state\ Probability | $P_{V\to V}$ | $P_{V\to H}$ | $P_{H\to V}$ | $P_{H\to H}$ |
|---|---|---|---|---|
| ($|g\rangle$) | 0.58 ±0.04 | 0.38 ±0.03 | 0.35 ±0.03 | 0.61 ±0.07 |
| ($|-\rangle$) | 0.10 ±0.07 | 0.98 ±0.04 | 0.93 ±0.03 | 0.07 ±0.07 |

**Table 1** Experiementally measured probabilities $P_{i\to j}$ where $i,j\in[H,V]$ for the two possible states of the control bit ($|g\rangle$ and $|-\rangle$).